\documentclass[twocolumn,twoside,slac_two]{revtex4}
\usepackage{graphicx}
\usepackage{fancyhdr}
\begin{document}

\title{Theoretical Overview of Hadronic Three-body $B$ Decays}

%

\author{Hai-Yang Cheng}
\affiliation{Institute of Physics, Academia Sinica, Taipei, Taiwan 115, Republic of China}

\begin{abstract}
Charmless 3-body decays of $B$ mesons are studied in a simple
model based on the factorization approach. We have
identified two distinct sources of nonresonant contributions: a small contribution from the tree transition and
a large source of the nonresonant signal in the matrix
elements of scalar densities, e.g., $\langle K\overline K|\bar
ss|0\rangle$, induced from the penguin transition. This explains the dominance of the nonresonant
background in $B\to KKK$ decays, the sizable nonresonant fraction in $K^-\pi^+\pi^-$ and $\overline
K^0\pi^+\pi^-$ modes and the smallness of nonresonant rates in $B\to \pi\pi\pi$ decays. The seemingly huge discrepancy between BaBar and Belle for the nonresonant contribution in the decay $B^-\to K^-\pi^+\pi^-$ is now relieved.
We have computed the resonant and nonresonant
contributions to charmless 3-body decays and determined the rates
for the quasi-two-body decays $B\to VP$ and $B\to SP$.
Time-dependent $CP$ asymmetries $\sin2\beta_{\rm eff}$ and $A_{CP}$
in $K^+K^-K_S,K_SK_SK_S,K_S\pi^+\pi^-$ and $K_S\pi^0\pi^0$ modes are estimated.
\end{abstract}

\maketitle

\thispagestyle{fancy}


\section{Introduction}

Recently many three-body $B$ decay modes, for example, $K^+\pi^+\pi^-$, $K^0\pi^+\pi^-$, $K^+\pi^-\pi^0$, $K^+K^+K^-$, $K^0K^+K^-$, $K_SK_SK^+$ and $\pi^+\pi^+\pi^-$, have been observed at $B$ factories with
branching ratios of order $10^{-5}$. The Dalitz plot analysis of
3-body $B$ decays provides a nice methodology for extracting
information on the unitarity triangle in the standard model. The
three-body meson decays are generally dominated by intermediate
vector and scalar resonances, namely, they proceed via
quasi-two-body decays containing a resonance state and a
pseudoscalar meson. Indeed, most of the quasi-two $B$ decays are
extracted from the analysis of three-body $B$ decays using the
Dalitz plot technique. Three-body hadronic $B$ decays involving a vector meson or charmed meson also have been observed at $B$ factories. In this talk I'll focus on  charmless 3-body $B$ decays.

Experimentally, there are two striking features for 3-body hadronic $B$ decays:

\begin{table}[b]
\caption{Fractions (\%) of nonresonant
contributions to various charmless three-body decays of $B$
mesons \cite{HFAG}. It will be explained below about the BaBar measurement of the nonresonant component in the $B^-\to K^-\pi^+\pi^-$ decay.}
\begin{tabular}{@{}lcc@{}} \toprule Decay & BaBar & Belle  \\
\colrule
$B^-\to K^+K^-K^-$ & $141\pm18$  & $74.8\pm3.6$ \\
$\overline B^0\to K^+K^-\overline K^0$ & $112\pm15$ &  \\
$\overline B^0\to\overline K^0\pi^+\pi^-$\hphantom{0} &  & $41.9\pm5.5$ \\
$B^-\to K^-\pi^+\pi^-$\hphantom{00} & \hphantom{0}$17.1^{+12.5}_{-~2.5}$ &
\hphantom{0}$34.0\pm2.9$ \\
$\overline B^0\to K^+\pi^-\pi^0$\hphantom{00} & \hphantom{0}$$ &
\hphantom{0}$15.6\pm7.7<25.7$ \\
$B^-\to\pi^+\pi^-\pi^-$\hphantom{00} & \hphantom{0}$13.6\pm6.1$ & \hphantom{0}  \\
\botrule
\end{tabular} \label{tab:BRexpt}
\end{table}

\vskip 0.2cm
\noindent {(i) large noresonant fractions in peguin-dominated modes}
\vskip 0.2cm
It is known that the nonresonant signal in charm decays is small,
less than 10\% \cite{PDG}. In the past few years, some of the
charmless $B$ to 3-body decay modes have been measured at $B$
factories and studied using the Dalitz plot analysis. We see from
Table \ref{tab:BRexpt} that the nonresonant fraction is about $\sim$ 90\%
in $B\to KKK$ decays, $\sim 17- 40\%$  in $B\to K\pi\pi$ decays (smaller in the $K\pi\pi^0$ decay), and $\sim$ 14\% in the $B\to\pi\pi\pi$
decay. Hence, the nonresonant 3-body decays play an essential
role in penguin-dominated $B$ decays. While this is a surprise in view of the rather
small nonresonant contributions in 3-body charm decays, it is not
entirely unexpected because the energy release scale in weak $B$
decays is of order 5 GeV, whereas the major resonances lie in the
energy region of 0.77 to 1.6 GeV. Consequently, it is likely that
3-body $B$ decays may receive sizable nonresonant contributions.
It is important to
understand and identify the underlying mechanism for nonresonant
decays.

Nonresonant amplitudes in charm decays are usually assumed to be uniform in phase space. However, this is no longer true in $B$ decays due to the large energy release in weak $B$ decays. While both BaBar and Belle have adopted the parametrization
 \begin{eqnarray} \label{eq:ANR}
A_{\rm NR}&=&(c_{12}e^{i\phi_{12}}e^{-\alpha
s_{12}^2}+c_{13}e^{i\phi_{13}}e^{-\alpha
s_{13}^2} \nonumber \\ &+& c_{23}e^{i\phi_{23}}e^{-\alpha s_{23}^2})(1+b_{\rm
NR}e^{i(\beta+\delta_{\rm NR})})
 \end{eqnarray}
to describe the non-resonant $B\to KKK$
amplitudes, they differ in the analysis of the nonresonant component in $B\to K\pi\pi$ decays. Belle still employed the above exponential parametrization to analyze the nonresonant contribution, but BaBar used the LASS parametrization  to describe the $K\pi$ $S$-wave and
the nonresonant component by a single amplitude suggested by the
LASS collaboration to describe the scalar amplitude in elastic
$K\pi$ scattering. Since the BaBar and Belle definitions of the $K_0^*(1430)$ and nonresonant differ, the branching fractions and phases are not directly comparable. We will come this point more in Sec. III.

Experimentally, it is hard to
measure the direct 3-body decays as the interference between
nonresonant and quasi-two-body amplitudes makes it difficult to
disentangle these two distinct contributions and extract the
nonresonant one.

\vskip 0.2cm
\noindent {(ii) New broad scalar resonances $f_X(1550)$ and $f_X(1300)$}
\vskip 0.2cm

A broad scalar resonance $f_X(1500)$ (or $X_0(1550)$ denoted by BaBar) has been seen in $B\to K^+K^+K^-$, $K^+K^-K_S$ and $K^+K^-\pi^+$ decays at energies around 1.5 GeV. However, it cannot be identified with the well known scaler meson $f_0(1500)$. This is because $f_0(1500)$ decays into $\pi^+\pi^-$ about five times more frequently than to $K^+K^-$. Identification of $f_X(1500)$ with $f_0(1500)$ will imply that the $K^+K^-$ peak at 1.5 GeV will be accompanied by a peak in $\pi^+\pi^-$, which is not seen experimentally. Hence, the nature of $f_X(1500)$ is not clear.

Moreover, there exists a production puzzle for $f_X(1500)$.
Both BaBar and Belle have seen a
large fraction from $f_X(1500)$ in the decay $B^-\to K^+K^-K^-$: $(121\pm19\pm6)\%$ by BaBar
\cite{BaBarKpKpKm} and $(63.4\pm6.9)\%$ by Belle
\cite{BelleKpKpKm}, whereas it is only about 4\% seen by BaBar in $B^0\to K^+K^-K_S$ \cite{BaBarKpKmK0}. The puzzle is that why $f_X(1500)$ behaves so dramatically different in charged and neutral $B$ decays to 3 kaons. It is not clear whether the large production of $f_X(1500)$ is a genuine effect or just a statistical fluctuation. Anyway, this issue should be clarified soon. Notice that Belle actually
found two solutions for the fraction of $f_X(1500)K^-$ in $B^-\to K^+K^+K^-$ \cite{BelleKpKpKm}:
$(63.4\pm6.9)\%$ and $(8.21\pm1.94)\%$. The first solution is
preferred by Belle. It is probably worth of re-examining the small solution.

\section{Three-body $B$ decays}
\begin{figure}[h]
\centering
\includegraphics[width=57mm]{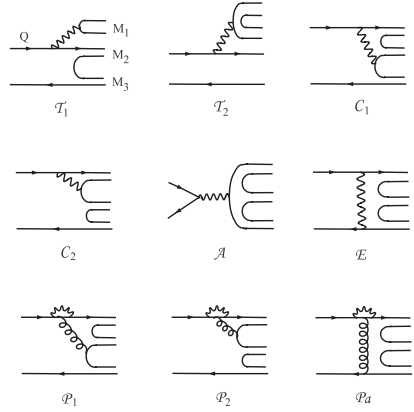}
\caption{Quark diagrams for the three-body decays of heavy
    mesons, where $Q$ denotes a heavy quark.} \label{QD}
\end{figure}

In analog to two-body decays of heavy mesons which can be analyzed using the model-independent quark diagrammatic approach,
three-body decays of the heavy mesons can be expressed in terms of
some quark-graph amplitudes \cite{CC90,CC87} (see Fig. \ref{QD}): ${\cal T}_1$ and ${\cal T}_2$,
the color-allowed external $W$-emission tree diagrams; ${\cal C}_1$ and
${\cal C}_2$, the color-suppressed internal $W$-emission diagrams; ${\cal E}$,
the $W$-exchange diagram; ${\cal A}$, the $W$-annihilation diagram;
${\cal P}_1$ and ${\cal P}_2$, the penguin diagrams, and ${\cal
P}_a$, the penguin-induced annihilation diagram. The quark-graph
amplitudes of various 3-body $B$ decays $B\to\pi h^+h^-$ and $B\to
Kh^+h^-$ are summarized in Table I of \cite{CYnonr}. As mentioned in \cite{CC90},
the use of the quark-diagram amplitudes for three-body decays are
in general momentum dependent. This means that unless its momentum
dependence is known, the quark-diagram amplitudes of direct 3-body
decays cannot be extracted from experiment without making further
assumptions. Moreover, the momentum dependence of each
quark-diagram amplitude varies from channel to channel.

We take the decay $\overline B^0\to K^+K^-\overline K^0$ as an
illustration. Under the factorization approach, its decay amplitude
consists of three distinct factorizable terms: (i) the transition
process induced by $b\to s$ penguins, $\langle \overline B^0\to
K^+\overline K^0\rangle\times \langle 0\to K^-\rangle$,  (ii) the
current-induced process through the tree $b\to u$ transition,
$\langle \overline B^0\to \overline K^0\rangle\times \langle 0\to
K^+K^-\rangle$,  and (iii) the annihilation process $\langle
\overline B^0\to 0\rangle\times \langle 0\to K^+K^-\overline
K^0\rangle$, where $\langle A\to B\rangle$ denotes a $A\to B$
transition matrix element.

\subsection{Nonresonant background}
For the transition process, the general expression of the
nonresonant contribution has the form
 \begin{eqnarray} \label{eq:AHMChPT}
 && \langle K^-(p_3)|(\bar s
 u)_{V-A}|0\rangle \langle\overline K {}^0 (p_1) K^+(p_2)|(\bar u b)_{V-A}|\overline B {}^0\rangle^{NR} \nonumber\\
 &&= -\frac{f_K}{2}\big[2 m_3^2 r
  +(m_B^2-s_{12}-m_3^2) \omega_+ \nonumber \\
 && \qquad +(s_{23}-s_{13}) \omega_-\big],
 \end{eqnarray}
where $(\bar q_1q_2)_{V-A}\equiv \bar
q_1\gamma_\mu(1-\gamma_5)q_2$. In principle, one can apply heavy
meson chiral perturbation theory (HMChPT) to evaluate the form
factors $r,~\omega_+$ and $\omega_-$ (for previous studies, see \cite{Fajfer}). However, this will lead to
too large decay rates in disagreement with experiment
\cite{Cheng:2002qu}. A direct calculation indicates that the
branching ratio of $\overline B^0\to K^+K^-\overline K^0$ arising
from the transition process alone is already at the level of
$77\times 10^{-6}$ which exceeds the measured total branching ratio
\cite{HFAG} of $25\times 10^{-6}$. The issue has to do with the
applicability of HMChPT. In order to apply this approach, two of
the final-state pseudoscalars ($K^+$ and $\overline K^0$ in this
example) have to be soft. The momentum of the soft pseudoscalar
should be smaller than the chiral symmetry breaking scale
$\Lambda_\chi$ of order $0.83-1.0$ GeV. For 3-body charmless $B$
decays, the available phase space where chiral perturbation theory
is applicable is only a small fraction of the whole Dalitz plot.
Therefore, it is not justified to apply chiral and heavy quark
symmetries to a certain kinematic region and then generalize it to
the region beyond its validity. If the soft meson result is assumed
to be the same in the whole Dalitz plot, the decay rate will be
greatly overestimated.

Recently we have proposed to parametrize the $b\to u$ trasnition-induced
nonresonant amplitude given by Eq. (\ref{eq:AHMChPT}) as
\cite{CCS3body}
 \begin{eqnarray} \label{eq:ADalitz}
  A_{\rm NR}=A_{\rm NR}^{\rm
  HMChPT}\,e^{-\alpha_{_{\rm NR}}
p_B\cdot(p_1+p_2)}e^{i\phi_{12}},
 \end{eqnarray}
so that the HMChPT results are recovered in the chiral limit
$p_1,~p_2\to 0$. That is, the nonresonant amplitude in the soft
meson region is described by HMChPT, but its energy dependence
beyond the chiral limit is governed by the exponential term
$e^{-\alpha_{_{\rm NR}} p_B\cdot(p_1+p_2)}$. The unknown parameter
$\alpha_{_{\rm NR}}$ can be determined from the data of the
tree-dominated decay $B^-\to\pi^+\pi^-\pi^-$.

In addition to the $b\to u$ tree transition, we need to consider the
nonresonant contributions to the $b\to s$ penguin amplitude
 \begin{eqnarray}
 A_1 &=& \langle \overline K {}^0|(\bar s b)_{V-A}|\overline B {}^0\rangle
  \langle K^+ K^-|(\bar u u)_{V-A}|0\rangle, \nonumber \\
 A_2 &=& \langle \overline K {}^0|\bar s b|\overline B {}^0\rangle
       \langle K^+ K^-|\bar s s|0\rangle.
 \end{eqnarray}
The 2-kaon creation matrix elements can be expressed in terms of
time-like kaon current form factors as
 \begin{eqnarray}\label{eq:KKweakff}
 \langle K^+(p_{K^+}) K^-(p_{K^-})|\bar q\gamma_\mu q|0\rangle
 &=& (p_{K^+}-p_{K^-})_\mu F^{K^+K^-}_q,
 \nonumber  \\
 \langle K^0(p_{K^0}) \overline K^0(p_{\bar K^0})|\bar q\gamma_\mu
 q|0\rangle
 &=& (p_{K^0}-p_{\bar K^0})_\mu F^{K^0\bar K^0}_q.
 \end{eqnarray}
The weak vector form factors $F^{K^+K^-}_q$ and $F^{K^0\bar K^0}_q$
can be related to the kaon electromagnetic (e.m.) form factors
$F^{K^+K^-}_{em}$ and $F^{K^0\bar K^0}_{em}$ for the charged and
neutral kaons, respectively. Phenomenologically, the e.m. form
factors receive resonant and nonresonant contributions
 \begin{eqnarray} \label{eq:KKemff}
 F^{K^+K^-}_{em} &=&  F_\rho+F_\omega+F_\phi+F_{NR}, \nonumber \\
 F^{K^0\bar K^0}_{em} &=&  -F_\rho+F_\omega+F_\phi+F_{NR}'.
 \end{eqnarray}
The resonant and nonresonant terms in Eq. (\ref{eq:KKemff}) can be
determined from a fit to the kaon e.m. data. The non-resonant
contribution to the matrix element $\langle K^+ K^-|\bar s
s|0\rangle$ is given by
 \begin{eqnarray} \label{eq:KKssme}
 && \langle K^+(p_2) K^-(p_3)|\bar s s|0\rangle^{NR}
 \equiv  f_s^{K^+K^-}(s_{23})\nonumber \\
 && =\frac{v}{3}(3 F_{NR}+2F'_{NR})+\sigma_{_{\rm NR}}
 e^{-\alpha\,s_{23}}.
 \end{eqnarray}
The nonresonant $\sigma_{_{\rm NR}}$ term is introduced for the
following reason. Although the nonresonant contributions to
$f_s^{KK}$ and $F_s^{KK}$ are related through the equation of
motion, the resonant ones are different and not related {\it a
priori}. As stressed in \cite{CCSKKK}, to apply the equation of
motion, the form factors should be away from the resonant region. In
the presence of the resonances, we thus need to introduce a
nonresonant $\sigma_{_{\rm NR}}$ term which can be constrained by
the measured $\overline B^0\to K_SK_SK_S$ rate and the $K^+K^-$ mass
spectrum \cite{CCS3body}.

\subsection{Resonant contributions}
Vector meson and scalar resonances contribute to the two-body matrix
elements $\langle P_1P_2|V_\mu|0\rangle$ and $\langle
P_1P_2|S|0\rangle$, respectively. They can also contribute to the
three-body matrix element $\langle P_1P_2|V_\mu-A_\mu|B\rangle$.
Resonant effects are described in terms of the usual Breit-Wigner
formalism. More precisely,
 \begin{eqnarray}
 && \langle K^+K^-|\bar q\gamma_\mu q|0\rangle^R = \sum_i\langle K^+K^-|V_i\rangle  \nonumber \\
&& \times {1\over m_{V_i}^2-s-im_{V_i}\Gamma_{V_i}}\langle V_i|\bar
q\gamma_\mu
q|0\rangle, \nonumber \\
&& \langle K^+K^-|\bar ss|0\rangle^R = \sum_i\langle K^+K^-|S_i\rangle \nonumber \\
&& \times {1\over m_{S_i}^2-s-im_{S_i}\Gamma_{S_i}}\langle S_i|\bar
ss|0\rangle,
 \end{eqnarray}
where $V_i=\phi,\rho,\omega,\cdots$ and
$S_i=f_0(980),f_0(1370),f_0(1500),\cdots$.
In this manner we are able to figure out the relevant resonances
which contribute to the 3-body decays of interest and compute the
rates of $B\to VP$ and $B\to SP$.

\begin{table*}[t]
\caption{Branching ratios (in units of $10^{-6}$) of resonant and
nonresonant (NR) contributions to $B^-\to K^-\pi^+\pi^-$.
Theoretical errors correspond to the uncertainties in (i)
$\alpha_{_{\rm NR}}$, (ii) $m_s$, $F^{BK}_0$ and $\sigma_{_{\rm
NR}}$, and (iii) $\gamma=(59\pm7)^\circ$. For the BaBar results, the branching fraction of $\overline K^{*0}_0(1430)\pi^-$ comes only from the Breit-Wigner component of the LASS parametrization, while the nonresonant contribution includes both the nonresonant part of the LASS shape and the phase-space nonresonant piece.}
\begin{ruledtabular}
\begin{tabular}{@{}l l l l@{}}
 Decay mode~~& BaBar \cite{BaBarKpipi} & Belle \cite{BelleKpipi} &
 Theory \cite{CCS3body}
\\ \colrule
 $\overline K^{*0}\pi^-$ & $7.2\pm0.4\pm0.7^{+0.3}_{-0.5}$
 & $6.45\pm0.43\pm0.48^{+0.25}_{-0.35}$ &  $3.0^{+0.0+0.8+0.0}_{-0.0-0.7-0.0}$ \\
 $\overline K^{*0}_0(1430)\pi^-$ & $32.0\pm1.2\pm2.7^{+9.1}_{-1.4}\pm5.2$ &
$32.0\pm1.0\pm2.4^{+1.1}_{-1.9}$ & $10.5^{+0.0+3.2+0.0}_{-0.0-2.7-0.1}$  \\
 $\rho^0K^-$ & $3.56\pm0.45\pm0.43^{+0.38}_{-0.15}$ &
$3.89\pm0.47\pm0.29^{+0.32}_{-0.29}$ & $1.3^{+0.0+1.9+0.1}_{-0.0-0.7-0.1}$ \\
 $f_0(980)K^-$ & $10.3\pm0.5\pm1.3^{+1.5}_{-0.4}$
 & $8.78\pm0.82\pm0.65^{+0.55}_{-1.64}$ & $7.7^{+0.0+0.4+0.1}_{-0.0-0.8-0.1}$  \\
NR & $9.3\pm1.0\pm1.2^{+6.7}_{-0.4}\pm1.2$ &
$16.9\pm1.3\pm1.3^{+1.1}_{-0.9}$ & $18.7^{+0.5+11.0+0.2}_{-0.6-~6.3-0.2}$ \\
\hline
 Total & $54.4\pm1.1\pm4.6$ & $48.8\pm1.1\pm3.6$ & $45.0^{+0.3+16.4+0.1}_{-0.4-10.5-0.1}$ \\
\end{tabular} \label{tab:Kpipi}
\end{ruledtabular}
\end{table*}

\begin{table*}[t]
\caption{Branching ratios (in units of $10^{-6}$) of resonant and
nonresonant (NR) contributions to $\overline B^0\to \overline K^0\pi^+\pi^-$.
}
\begin{ruledtabular} \label{tab:K0pipi}
\begin{tabular}{l l l }
 Decay mode~~ &  Belle \cite{BelleK0pipi} & Theory \cite{CCS3body} \\ \hline
 $K^{*-}\pi^+$ & $5.6\pm0.7\pm0.5^{+0.4}_{-0.3}$ &  $2.1^{+0.0+0.5+0.3}_{-0.0-0.5-0.3}$ \\
 $K^{*-}_0(1430)\pi^+$  & $30.8\pm2.4\pm2.4^{+0.8}_{-3.0}$ & $10.1^{+0.0+2.9+0.1}_{-0.0-2.5-0.2}$ \\
 $\rho^0\overline K^0$  & $6.1\pm1.0\pm0.5^{+1.0}_{-1.1}$ & $2.0^{+0.0+1.9+0.1}_{-0.0-0.9-0.1}$ \\
 $f_0(980)\overline K^0$ & $7.6\pm1.7\pm0.7^{+0.5}_{-0.7}$ & $7.7^{+0.0+0.4+0.0}_{-0.0-0.7-0.0}$ \\
 NR  & $19.9\pm2.5\pm1.6^{+0.7}_{-1.2}$ & $15.6^{+0.1+8.3+0.0}_{-0.1-4.9-0.0}$ \\ \hline
 Total  & $47.5\pm2.4\pm3.7$ & $42.0^{+0.3+15.7+0.0}_{-0.2-10.8-0.0}$ \\
\end{tabular}
\end{ruledtabular}
\end{table*}

\begin{table*}[t]
\caption{Branching ratios (in units of $10^{-6}$) of resonant and
nonresonant (NR) contributions to $\overline B^0\to K^-\pi^+\pi^0$. For the BaBar measurement, the resonance $K_0^*(1430)$ is replaced by the $S$-wave $K\pi$ state, namely, $(K\pi)_0^*$.} \label{tab:Kmpippi0}
\begin{ruledtabular}
\begin{tabular}{ l l l l}
 Decay mode~~ & BaBar \cite{BaBarKppimpi0} & Belle \cite{BelleKppimpi0} & Theory \cite{CCS3body} \\ \colrule
 $K^{*-}\pi^+$ & $4.2^{+0.9}_{-0.5}\pm0.3$  & $4.9^{+1.5+0.5+0.8}_{-1.5-0.3-0.3}$ &   $1.0^{+0.0+0.3+0.1}_{-0.0-0.3-0.1}$ \\
 $\overline K^{*0}\pi^0$ & $2.4\pm0.5\pm0.3$  & $<2.3$ &  $1.0^{+0.0+0.3+0.2}_{-0.0-0.2-0.1}$  \\
 $K^{*-}_0(1430)\pi^+$ & $9.4^{+1.1+1.4}_{-1.3-1.1}\pm1.8$  &  $5.1\pm1.5^{+0.6}_{-0.7}$ & $5.0^{+0.0+1.5+0.1}_{-0.0-1.3-0.1}$ \\
 $\overline K^{*0}_0(1430)\pi^0$ & $8.7^{+1.1+1.8}_{-0.9-1.3}\pm2.2$ &  $6.1^{+1.6+0.5}_{-1.5-0.6}$ & $4.2^{+0.0+1.4+0.0}_{-0.0-1.2-0.0}$ \\
 $\rho^+K^-$ & $8.0^{+0.8}_{-1.3}\pm0.6$ & $15.1^{+3.4+1.4+2.0}_{-3.3-1.5-2.1}$ & $2.5^{+0.0+3.6+0.2}_{-0.0-1.4-0.2}$ \\
 NR &  & $5.7^{+2.7+0.5}_{-2.5-0.4}<9.4$ & $9.6^{+0.3+6.6+0.0}_{-0.2-3.5-0.0}$ \\ \colrule
 Total & $35.7^{+2.6}_{-1.5}\pm2.2$ & $36.6^{+4.2}_{-4.1}\pm3.0$ & $28.9^{+0.2+16.1+0.2}_{-0.2-~9.4-0.2}$ \\
\end{tabular}
\end{ruledtabular}
\end{table*}

\begin{table*}[t]
\caption{Same as Table \ref{tab:Kpipi} except for  $B^-\to
\pi^+\pi^-\pi^-$. The nonresonant background is used as an input to
fix the parameter $\alpha_{_{\rm NR}}$ defined in Eq.
(\ref{eq:ADalitz}). }
\begin{ruledtabular} \label{tab:Bpipipi}
\begin{tabular}{l l l }
 Decay mode~~ &  BaBar \cite{BaBarpipipi} & Theory \cite{CCS3body} \\ \hline
 $\rho^0\pi^-$ & $8.8\pm1.0\pm0.6^{+0.1}_{-0.7}$ &  $7.7^{+0.0+1.7+0.3}_{-0.0-1.6-0.2}$ \\
 $f_0(980)\pi^-$ & $1.2\pm0.6\pm0.1\pm0.4<3.0$ & $0.39^{+0.00+0.01+0.03}_{-0.00-0.01-0.02}$ \\
NR  & $2.3\pm0.9\pm0.3\pm0.4<4.6$ & input \\ \hline
 Total  & $16.2\pm1.2\pm0.9$ & $12.0^{+1.1+2.0+0.4}_{-1.2-1.8-0.3}$ \\
\end{tabular}
\end{ruledtabular}
\end{table*}

\section{Penguin-dominated $B\to KKK$ and $B\to K\pi\pi$ decays}

As mentioned in the previous section, we employ the decays
$\overline B^0\to K^+K^-\overline K^0$ and $K_SK_SK_S$ to fix the
nonresonant parameter $\sigma_{_{\rm NR}}$ to be
 \begin{eqnarray} \label{eq:sigma}
  \sigma_{_{\rm NR}}= e^{i\pi/4}\left(3.36^{+1.12}_{-0.96}\right)\,{\rm GeV}.
 \end{eqnarray}
It turns out that the nonresonant contribution arises dominantly
from the transition process (88\%) via the scalar-density-induced
vacuum to $K\bar K$ transition, namely, $\langle K^+K^-|\bar
ss|0\rangle$, and slightly from the current-induced process (3\%).
Physically, this is because the decay $B\to KKK$ is dominated by the
$b\to s$ penguin transition. The nonresonant background in $B\to KK$
transition does not suffice to account for the experimental
observation that the penguin-dominated decay $B\to KKK$ is dominated
by the nonresonant contributions. This implies that the two-body
matrix element e.g. $\langle K\overline K|\bar ss|0\rangle$ induced
by the scalar density should have a large nonresonant component.

We have considered other $B\to KKK$ decays such as $B^-\to
K^+K^-K^-$ and $B^-\to K^-K_SK_S$ and found that they are also
dominated by the nonresonant contributions. Our predicted branching
ratio ${\cal B}(B^-\to K^+K^-K^-)_{_{\rm
NR}}=(25.3^{+4.9}_{-4.5})\times 10^{-6}$ \cite{CCS3body} is in good agreement with
the Belle measurement of $(24.0^{+3.0}_{-6.2})\times 10^{-6}$, but a factor of 2 smaller than the BaBar result of $(50\pm6\pm4)\times
10^{-6}$ \cite{BaBarKpKpKm}.

The resonant and nonresonant contributions to the decay $B^-\to
K^-\pi^+\pi^-$ are shown in Table \ref{tab:Kpipi}. We see that the
calculated $K^*\pi$ and $\rho K$ rates are smaller than the data by
a factor of $2\sim 3$. This seems to be a generic feature of the
factorization approach such as QCD factorization where the predicted
penguin-dominated $VP$ rates are too small compared to experiment.
We shall return back to this point later.

At first sight, it appears that the nonresonant branching ratio $(2.4\pm0.5\pm1.3^{+0.3}_{-0.8})\times 10^{-6}$ in $B^-\to K^-\pi^+\pi^-$ obtained by BaBar \cite{BaBarKpipi} is much smaller than the one $(16.9\pm1.3\pm1.3^{+1.1}_{-0.9})\times 10^{-6}$ measured by Belle \cite{BelleKpipi}. However as mentioned in the Introduction, since the BaBar and Belle definitions of the $K_0^*(1430)$ and nonresonant differ, it does not make sense to compare the branching fractions and phases directly.
While Belle
\cite{BelleKpipi} employed the exponential parametrization Eq. (\ref{eq:ANR}) to
describe the nonresonant contribution, BaBar \cite{BaBarKpipi} used
the LASS parametrization to describe the $K\pi$ $S$-wave and the
nonresonant component by a single amplitude suggested by the LASS
collaboration
 \begin{eqnarray}
 {\cal M} &=& {m_{K\pi}\over q\cot \delta_B-iq} \nonumber \\
 &+& e^{2i\delta_B}{m_0\Gamma_0{m_0\over q_0}\over (m_0^2-m_{K\pi}^2)-im_0\Gamma_0{q\over m_{K\pi}}{m_0\over q_0}},
 \end{eqnarray}
where $\cot\delta_B={1\over aq}+{1\over 2}rq$.
Since the LASS parametrization is valid (experimentally confirmed) up
to the $K\pi$ invariant mass of order 1.8 GeV, BaBar introduced a phase-space
 nonresonant component to describe an excess of signal events at higher $K\pi$ invariant mass. Hence, the BaBar definition for the $K^*_0(1430)$ includes an effective range term to account for the low
$K\pi$ S-wave while for the Belle parameterization, this component is absorbed into the nonresonant piece. To stress once again, the result ${\cal B}(B^-\to K^-\pi^+\pi^-)_{_{\rm NR}}=(2.4\pm0.5\pm1.3^{+0.3}_{-0.8})\times 10^{-6}$ cited by BaBar is solely due to the phase-space nonresonant piece.

From the above discussion, it is clear that part of the LASS shape is really nonresonant which
has a substantial mixing with $K^*_0(1430)$. In principle, this should be added to the phase-space nonresonant piece to get the total nonresonant contribution.  Once this is done, it is
possible that BaBar and Belle might agree with each other. Indeed, very recently BaBar have carried out this task \cite{BaBarKpipi}. By combining coherently the nonresonant part of the LASS parametrization and the phase-space nonresonant, BaBar found the total nonresonant branching fraction to be  $(9.3\pm1.0\pm1.2^{+6.7}_{-0.4}\pm1.2)\times 10^{-6}$ with the fit fraction being $(17.1\pm1.7\pm1.6^{+12.3}_{-~0.8})\%$ \cite{Gershon}.
We see from  Table \ref{tab:Kpipi} that the BaBar result is now
consistent with Belle within errors, though the agreement is not perfect as BaBar and Belle have different models for the nonresonant $K\pi$ mass. Likewise, the BaBar branching fraction $(24.5\pm0.9\pm2.1^{+7.0}_{-1.1})\times 10^{-6}$ for $B^-\to K_0^*(1430)\pi^-$ cited in Table II of \cite{BaBarKpipi} includes an effective range nonresonant component. In order to compare with the Belle result determined from the Breit-Wigner parametrization,
it would be more appropriate to consider the Breit-Wigner component only of the LASS parametrizaion. The result is ${\cal B}(B^-\to K_0^*(1430)\pi^-)=(32.0\pm1.2\pm2.7^{+9.1}_{-1.4}\pm5.2)\times 10^{-6}$ which is now in good agreement with the Belle measurement (see Table \ref{tab:Kpipi}).

From Table \ref{tab:Kpipi} we see that our predicted nonresonant
rates are consistent with the Belle and BaBar measurements within errors. The reason for the large nonresonant rates in the
$K^-\pi^+\pi^-$ mode is that under SU(3) flavor symmetry, we have
the relation $\langle K\pi|\bar sq|0\rangle^{NR}=\langle K\bar
K|\bar ss|0\rangle^{NR}$.
Hence, the nonresonant rates in the $K^-\pi^+\pi^-$ (Table \ref{tab:Kpipi})and $\overline
K^0\pi^+\pi^-$ (Table \ref{tab:K0pipi}) modes should be similar to that in $K^+K^-\overline
K^0$ or $K^+K^-K^-$. Since the $KKK$ channel receives resonant
contributions only from $\phi$ and $f_{0}$ mesons, while $K^*,
K^*_{0},\rho,f_{0}$ resonances contribute to $K\pi\pi$ modes, this
explains why the nonresonant fraction is of order 90\% in the former
and becomes of order 40\% in the latter. It is interesting to notice
that, based on a simple fragmentation model and SU(3) symmetry,
Gronau and Rosner \cite{Gronau3body} also found  a large nonresonant
background in $K^-\pi^+\pi^-$ and $\overline K^0\pi^+\pi^-$.

Recently, BaBar has reported a new Dalitz-plot analysis
of the decay $\overline B^0\to K^-\pi^+\pi^0$  \cite{BaBarKppimpi0} (see Table \ref{tab:Kmpippi0}). Just as the $K^-\pi^+\pi^-$ mode, the reported nonresonant branching fraction ${\cal B}(\overline B^0\to K^-\pi^+\pi^0)_{\rm nr}=(4.4\pm0.9\pm0.5)\times 10^{-6}$ by BaBar is only the phase-space part of nonresonant contributions. To get the total nonresonant rate, it is necessary to add the nonresonant component of the LASS parametrization to the phase-space piece. When this is done, it will be interesting to compare the measured nonresonant  branching fraction with our prediction ${\cal B}(\overline B^0\to K^-\pi^+\pi^0)_{\rm nr}=(9.6^{+0.3+6.6}_{-0.2-3.5})\times 10^{-6}$. It should be stressed that the measured partial rates for $\overline B^0\to (K\pi)_0^{*-}\pi^+$ and $(K\pi)^{*0}_0\pi^0$ by BaBar (see Table \ref{tab:Kmpippi0}) include an effective range $K\pi$ nonresonant component. Hence, it is not pertinent to compared them directly with the respective Belle measurements.

\section{Tree-dominated $B\to \pi\pi\pi,KK\pi$ modes}
The $B\to \pi\pi\pi$ mode receives nonresonant contributions mostly
from the $b\to u$ transition as the nonresonant contribution in the
penguin matrix element $\langle\pi^+\pi^-|\bar dd|0\rangle$ is
suppressed by the smallness of penguin Wilson coefficients $a_6$ and
$a_8$. Hence,  the measurement of the nonresonant
contribution in this decay can be used to constrain the
nonresonant parameter $\alpha_{_{\rm NR}}$ in Eq.
(\ref{eq:ADalitz}).

Note that while $B^-\to \pi^+\pi^-\pi^-$ is dominated by the
$\rho^0$ pole (Table \ref{tab:Bpipipi}), the decay $\overline B^0\to \pi^+\pi^-\pi^0$ receives
$\rho^\pm$ and $\rho^0$ contributions. As a consequence, the
$\pi^+\pi^-\pi^0$ mode has a rate larger than $\pi^+\pi^-\pi^-$ even
though the former involves a $\pi^0$ in the final state. We predict that ${\cal B}(B^0\to\pi^+\pi^-\pi^0)\approx 26\times 10^{-6}$ \cite{CCS3body}.

Among the 3-body decays we have studied, the decay $B^-\to
K^+K^-\pi^-$ dominated by $b\to u$ tree transition and $b\to d$
penguin transition has the smallest branching ratio of order
$4\times 10^{-6}$. BaBar \cite{BaBar:KKpi} has recently reported
the observation of the decay $B^+\to K^+K^-\pi^+$ with the
branching ratio $(5.0\pm0.5\pm0.5)\times 10^{-6}$. Our prediction
for this mode, $(4.0^{+0.5+0.7+0.3}_{-0.6-0.5-0.3})\times 10^{-6}$, is in accordance with experiment.

\section{Quasi-two-body $B$ decays}
It is known that in the narrow width approximation, the 3-body
decay rate obeys the factorization relation
 \begin{eqnarray} \label{eq:fact}
 \Gamma(B\to RP\to P_1P_2P)=\Gamma(B\to RP){\cal B}(R\to P_1P_2), \nonumber\\
 \end{eqnarray}
with $R$ being a vector meson or a scalar resonance.
Using the experimental information on ${\cal B}(R\to
h_2h_3)$ \cite{PDG}
  \begin{eqnarray}
{\cal B}(K^{*0}\to K^+\pi^-)&=&{\cal B}(K^{*+}\to K^0\pi^+) \nonumber \\ &=&2{\cal B}(K^{*+}\to
K^+\pi^0)={2\over 3}, \nonumber \\
{\cal B}(K_0^{*0}(1430)\to K^+\pi^-)&=&2{\cal B}(K_0^{*+}(1430)\to
 K^+\pi^0)\nonumber \\ &=&{2\over 3}(0.93\pm0.10), \nonumber \\ {\cal B}(\phi\to
 K^+K^-)&=&0.492\pm0.006\,,
  \end{eqnarray}
we have extracted the branching ratios of $B\to VP$ and $B\to SP$.
The results are summarized in Table \ref{tab:BR2body}.
 The
predicted $\rho\pi,~f_0(980)K$ and $f_0(980)\pi$ rates are in
agreement with the data, while the calculated $\phi K,~K^*\pi,~\rho
K$ and $K_0^*(1430)\pi$ are in general too small compared to
experiment. The fact that this work and QCDF lead to too small rates
for $\phi K,~K^*\pi,~\rho K$ and $K_0^*(1430)\pi$ may imply the
importance of power corrections due to the non-vanishing $\rho_A$
and $\rho_H$ parameters arising from weak annihilation and hard
spectator interactions, respectively, which are used to parametrize
the endpoint divergences, or due to possible final-state
rescattering effects from charm intermediate states \cite{CCSfsi}.
However, this is beyond the scope of the present work.

\begin{table*}[!]
\caption{Branching ratios of quasi-two-body decays $B\to VP$ and $B\to
SP$ obtained from the studies of three-body decays based on the
factorization approach \cite{CCS3body}.  Theoretical uncertainties have been
added in quadrature. QCD factorization predictions taken from
\cite{BN} for $VP$ modes and from \cite{CCY}
 for $SP$ channels are shown here for comparison. The assumption of
${\cal B}(f_0(980)\to\pi^+\pi^-)=0.50$ has been made for the QCDF calculation. Experimental results are taken from
\cite{HFAG}. Note that the BaBar results for $\bar K_0^{*0}(1430)\pi^0$ and $K_0^{*-}(1430)\pi^+$ are obtained by neglecting nonresonant contributions to $(K\pi)_0^*\pi$ \cite{BaBarKppimpi0} and hence may not be appropriate to compare with Belle directly. }
\begin{ruledtabular} \label{tab:BR2body}
\begin{tabular}{l c c | c c}
 Decay mode~~ &  BaBar & Belle  & QCDF & Theory \cite{CCS3body}  \\ \colrule
 $\phi K^0$ & $8.4^{+1.5}_{-1.3}\pm0.5$   &
 $9.0^{+2.2}_{-1.8}\pm0.7$   & $4.1^{+0.4+1.7+1.8+10.6}_{-0.4-1.6-1.9-~3.0}$ & $5.3^{+1.0}_{-0.9}$ \\
 $\phi K^-$ & $8.4\pm0.7\pm0.7$ & $9.60\pm0.92^{+1.05}_{-0.84}$ &
 $4.5^{+0.5+1.8+1.9+11.8}_{-0.4-1.7-2.1-~3.3}$ & $5.9^{+1.1}_{-1.0}$ \\
 $\overline K^{*0}\pi^-$ & $10.8\pm0.6^{+1.1}_{-1.3}$ &
 $9.7\pm0.6^{+0.8}_{-0.9}$ & $3.6^{+0.4+1.5+1.2+7.7}_{-0.3-1.4-1.2-2.3}$ & $4.4^{+1.1}_{-1.0}$ \\
 $\overline K^{*0}\pi^0$ & $3.6\pm0.7\pm0.4$ & $<3.5$ & $0.7^{+0.1+0.5+0.3+2.6}_{-0.1-0.4-0.3-0.5}$
 & $1.5^{+0.5}_{-0.4}$ \\
 $K^{*-}\pi^+$ & $11.7^{+1.3}_{-1.2}$ & $8.4\pm1.1^{+0.9}_{-0.8}$ &
 $3.3^{+1.4+1.3+0.8+6.2}_{-1.2-1.2-0.8-1.6}$ & $3.1^{+0.9}_{-0.9}$ \\
 $K^{*-}\pi^0$ & $6.9\pm2.0\pm1.3$  & & $3.3^{+1.1+1.0+0.6+4.4}_{-1.0-0.9-0.6-1.4}$ &
 $2.2^{+0.6}_{-0.5}$ \\
 $K^{*0}K^-$ & $<1.1$ & &
 $0.30^{+0.11+0.12+0.09+0.57}_{-0.09-0.10-0.09-0.19}$ & $0.35^{+0.06}_{-0.06}$ \\
 $\rho^0K^-$ & $3.56\pm0.45^{+0.57}_{-0.46}$ &
 $3.89\pm0.47^{+0.43}_{-0.41}$ & $2.6^{+0.9+3.1+0.8+4.3}_{-0.9-1.4-0.6-1.2}$ & $1.3^{+1.9}_{-0.7}$  \\
 $\rho^0\overline K^0$ & $4.9\pm0.8\pm0.9$ & $6.1\pm1.0\pm1.1$
 & $4.6^{+0.5+4.0+0.7+6.1}_{-0.5-2.1-0.7-2.1}$ & $2.0^{+1.9}_{-0.9}$ \\
 $\rho^+K^-$ & $8.0^{+0.8}_{-1.3}\pm0.6$ & $15.1^{+3.4+2.4}_{-3.3-2.6}$ &
 $7.4^{+1.8+7.1+1.2+10.7}_{-1.9-3.6-1.1-~3.5}$  & $2.5^{+3.6}_{-1.4}$ \\
 $\rho^-\overline K^0$ & $8.0^{+1.4}_{-1.3}\pm0.6$  & &
 $5.8^{+0.6+7.0+1.5+10.3}_{-0.6-3.3-1.3-~3.2}$ & $1.3^{+3.0}_{-0.9}$ \\
 $\rho^0\pi^-$ & $8.8\pm1.0^{+0.6}_{-0.9}$ &
 $8.0^{+2.3}_{-2.0}\pm0.7$   & $11.9^{+6.3+3.6+2.5+1.3}_{-5.0-3.1-1.2-1.1}$  & $7.7^{+1.7}_{-1.6}$ \\
 $\rho^-\pi^+$ & & &
 $21.2^{+10.3+8.7+1.3+2.0}_{-~8.4-7.2-2.3-1.6}$ & $15.5^{+4.0}_{-3.5}$ \\
 $\rho^+\pi^-$ & & & $15.4^{+8.0+5.5+0.7+1.9}_{-6.4-4.7-1.3-1.3}$ & $8.5^{+1.1}_{-1.0}$  \\
 $\rho^0\pi^0$ & $1.4\pm0.6\pm0.3$ & $3.0\pm0.5\pm0.7$ &
 $0.4^{+0.2+0.2+0.9+0.5}_{-0.2-0.1-0.3-0.3}$ & $1.0^{+0.3}_{-0.2}$ \\
 $f_0(980)K^0;f_0\to \pi^+\pi^-$ & $5.5\pm0.7\pm0.6$ & $7.6\pm1.7^{+0.8}_{-0.9}$ &
 $6.7^{+0.1+2.1+2.3}_{-0.1-1.5-1.1}$   & $7.7^{+0.4}_{-0.7}$ \\
 $f_0(980)K^-;f_0\to \pi^+\pi^-$ & $9.3\pm1.0^{+0.6}_{-0.9}$ &
 $8.8\pm0.8^{+0.9}_{-1.8}$ & $7.8^{+0.2+2.3+2.7}_{-0.2-1.6-1.2}$  & $7.7^{+0.4}_{-0.8}$ \\
 $f_0(980)K^0;f_0\to K^+K^-$ & $5.3\pm2.2$  &  & & $5.8^{+0.1}_{-0.5}$ \\
 $f_0(980)K^-;f_0\to K^+K^-$ & $6.5\pm2.5\pm1.6$ & $<2.9$ &  & $7.0^{+0.4}_{-0.7}$ \\
 $f_0(980)\pi^-;f_0\to \pi^+\pi^-$ & $<3.0$ & & $0.5^{+0.0+0.2+0.1}_{-0.0-0.1-0.0}$
 & $0.39^{+0.03}_{-0.02}$ \\
 $f_0(980)\pi^-;f_0\to K^+K^-$ & & & & $0.50^{+0.06}_{-0.04}$ \\
 $f_0(980)\pi^0;f_0\to \pi^+\pi^-$ & & &
 $0.02^{+0.01+0.02+0.04}_{-0.01-0.00-0.01}$  & $0.010^{+0.003}_{-0.002}$ \\
 $\overline K^{*0}_0(1430)\pi^-$ & $32.0\pm1.2^{+10.8}_{-~6.0}$ &
 $51.6\pm1.7^{+7.0}_{-7.4}$ & $11.0^{+10.3+7.5+49.9}_{-~6.0-3.5-10.1}$ & $16.9^{+5.2}_{-4.4}$ \\
 $\overline K^{*0}_0(1430)\pi^0$ & $13.1^{+1.6+2.7}_{-1.5-1.9}\pm3.6$ & $9.8\pm2.5\pm0.9$ &
 $6.4^{+5.4+2.2+26.1}_{-3.3-2.1-~5.7}$ & $6.8^{+2.3}_{-1.9}$ \\
 $K^{*-}_0(1430)\pi^+$ & $28.2^{+3.3+4.3}_{-4.1-3.3}\pm5.2$ & $49.7\pm3.8^{+4.0}_{-6.1}$
 & $11.3^{+9.4+3.7+45.8}_{-5.8-3.7-~9.9}$  & $16.2^{+4.7}_{-4.0}$ \\
 $K^{*-}_0(1430)\pi^0$ & & & $5.3^{+4.7+1.6+22.3}_{-2.8-1.7-~4.7}$
 &  $8.9^{+2.6}_{-2.2}$ \\
 $K^{*0}_0(1430)K^-$ & $<2.2$  & & & $1.3^{+0.3}_{-0.3}$ \\
\end{tabular}
\end{ruledtabular}
\end{table*}

\section{Time-dependent $CP$ asymmetries}
The penguin-induced three-body decays $B^0\to K^+K^-K_S$ and
$K_SK_SK_S$ deserve special attention as the current measurements of
the deviation of $\sin 2\beta_{\rm eff}$ in $KKK$ modes from $\sin 2
\beta_{J/\psi K_S}$ may indicate New Physics in $b\to s$
penguin-induced modes.  It is of great importance to examine and
estimate how much of the deviation of $\sin 2\beta_{\rm eff}$ is
allowed in the SM. Owing to the presence of color-allowed tree
contributions in $B^0\to K^+K^-K_{S}$, this mode is subject to a
potentially significant tree pollution and the deviation of the
mixing-induced $CP$ asymmetry from that measured in $B\to J/\psi
K_S$ could be as large as ${\cal O}(0.10)$. Since the tree amplitude
is tied to the nonresonant background, it is very important to
understand the nonresonant contributions in order to have a reliable
estimate of $\sin 2\beta_{\rm eff}$ in $KKK$ modes.

\begin{table*}[t]
\caption{Mixing-induced and direct $CP$ asymmetries for various
charmless 3-body $B$ decays. Experimental results are taken from
\cite{HFAG}.}
\begin{ruledtabular}
\begin{tabular}{@{}l r r r | r r @{} }
 Decay & $\sin 2\beta_{\rm eff}$ & $\Delta\sin 2\beta_{\rm eff}$ & Expt &
 $A_f(\%)$ & Expt \\ \colrule
$K^+K^-K_S$ & $0.728^{+0.001+0.002+0.009}_{-0.002-0.001-0.020}$ &
$0.041^{+0.028}_{-0.033}$ & $0.05\pm0.11$ & $-4.63^{+1.35+0.53+0.40}_{-1.01-0.54-0.34}$ & $-7\pm8$  \\
$K_SK_SK_S$ & $0.719^{+0.000+0.000+0.008}_{-0.000-0.000-0.019}$ &
$0.039^{+0.027}_{-0.032}$ & $-0.10\pm0.20$ & $0.69^{+0.01+0.01+0.05}_{-0.01-0.03-0.07}$ & $14\pm15$ \\
$K_S\pi^0\pi^0$ & $0.729^{+0.000+0.001+0.009}_{-0.000-0.001-0.020}$
& $0.049^{+0.027}_{-0.032}$ & $-1.20\pm0.41$ & $0.28^{+0.09+0.07+0.02}_{-0.06-0.06-0.02}$ & $-18\pm22$ \\
$K_S\pi^+\pi^-$ & $0.718^{+0.001+0.017+0.008}_{-0.001-0.007-0.018}$
& $0.038^{+0.031}_{-0.032}$ & & $4.94^{+0.03+0.03+0.32}_{-0.02-0.05-0.40}$ & \\
\end{tabular} \label{tab:CP}
\end{ruledtabular}
\end{table*}

The deviation of the mixing-induced $CP$ asymmetry in $B^0\to
K^+K^-K_S$, $K_SK_SK_S$, $K_S\pi^+\pi^-$ and $K_S\pi^0\pi^0$ from
that measured in $B\to \phi_{c\bar c}K_S$, i.e. $\sin 2
\beta_{\phi_{c\bar c}K_S}=0.681\pm0.025$ \cite{HFAG}, namely,
$\Delta \sin 2\beta_{\rm eff}\equiv \sin 2\beta_{\rm eff}-\sin 2
\beta_{\phi_{c\bar c}K_S}$, is shown in Table \ref{tab:CP}. Our
calculation indicates the deviation of the mixing-induced $CP$
asymmetry in $\overline B^0\to K^+K^-K_{S}$ from that measured in
$\overline B^0\to \phi_{c\bar c}K_S$ is very similar to that of the
$K_SK_SK_S$ mode as the tree pollution effect in the former is
somewhat washed out. Nevertheless, direct $CP$ asymmetry of the
former, being of order $-4\%$, is more prominent than the latter.

\section{Conclusions}
It is important to account for the large nonresonant amplitudes in the study of charmless 3-body baryonic $B$ decays.
We have
identified two distinct sources of nonresonant contributions: a small contribution from the tree transition and
a large source of the nonresonant signal in the matrix
elements of scalar densities, e.g. $\langle K\overline K|\bar
ss|0\rangle$, induced from the penguin transition. This explains the dominance of the nonresonant
background in $B\to KKK$ decays, the sizable nonresonant fraction in $K^-\pi^+\pi^-$ and $\overline
K^0\pi^+\pi^-$ modes and the smallness of nonresonant rates in $B\to \pi\pi\pi$ decays. The seemingly huge discrepancy between BaBar and Belle for the nonresonant contribution in the decay $B^-\to K^-\pi^+\pi^-$ is now relieved. Since penguin contributions to charm decays are GIM suppressed, hence nonresonant signals in $D$ decays are always small.

We have computed the resonant and nonresonant
contributions to charmless 3-body decays and determined the rates
for the quasi-two-body decays $B\to VP$ and $B\to SP$.
Time-dependent $CP$ asymmetries $\sin2\beta_{\rm eff}$ and $A_{CP}$
in $K^+K^-K_S,K_SK_SK_S,K_S\pi^+\pi^-$ and $K_S\pi^0\pi^0$ modes are estimated. Since we have a realistic model for resonant and nonresonant contributions, our estimation of $\sin2\beta_{\rm eff}$ for 3-body $B$ decays should be more reliable and trustworthy.

\section*{Acknowledgments}
I'm grateful to Chun-Khiang Chua and Amarjit Soni for fruitful
collaboration, to Tim Gershon and Jim Smith   for discussions and to Paoti Chang and Hsiang-nan Li for organizing this stimulating conference.



\begin{thebibliography}{00}
\bibitem{Gershon} I would like to thank Tim Gershon for providing me this number.
\end{thebibliography}
\end{document}